\documentclass[
showpacs,
floatfix,
aps,
prb,
amsmath,
twocolumn,
superscriptaddress, 
tightenlines,
groupaddress,
]{revtex4-1}

\usepackage{graphicx}
\usepackage{dcolumn}
\usepackage{amsmath}
\usepackage{amsfonts}
\usepackage{bm}
\usepackage{epsfig}

\newcommand{\be}{\begin{equation}}
\newcommand{\ee}{\end{equation}}
\newcommand{\bea}{\begin{eqnarray}}
\newcommand{\eea}{\end{eqnarray}}

\def\lp{\left (}
\def\rp{\right )}

\def\pc{\mathcal{P}_\omega}

\def\Edc{\mathcal{E}_j}
\def\W{\mathcal{W}_j}

\def\tst{\tau_\star}
\def\eac{\epsilon_{\omega}}
\def\edc{\epsilon_{j}}

\def\aac{A}
\def\adc{\alpha}

\def\oc{\omega_{\mbox{\scriptsize {c}}}}

\def\rc{R_{\mbox{\scriptsize {c}}}}

\def\tq{\tau_{\mbox{\scriptsize {q}}}}

\def\ttr{\tau_{\mbox{\scriptsize {tr}}}}

\def\tsh{\tau_{\mbox{\scriptsize {sh}}}}

\def\ttr{\tau}
\def\tin{\tau_{\mbox{\scriptsize {in}}}}

\newcommand{\req}[1]{Eq.\,(\ref{#1})}
\newcommand{\reqs}[2]{Eqs.\,(\ref{#1}) and (\ref{#2})}
\newcommand{\rfig}[1]{Fig.\,\ref{#1}}
\newcommand{\rref}[1]{Ref.\,\onlinecite{#1}}
\newcommand{\rrefs}[2]{Refs.\,\onlinecite{#1},\,\onlinecite{#2}}

\def\mob{$\mu \approx 8.9 \times 10^6$ cm$^2$/Vs}
\def\den{$n_e \approx 3.95 \times 10^{11}$ cm$^{-2}~$}

\begin{document}

\title{
Nonlinear response in overlapping and separated Landau levels of GaAs quantum wells
}

\author{A.\,T. Hatke}
\affiliation{School of Physics and Astronomy, University of Minnesota, Minneapolis, Minnesota 55455, USA}

\author{M.\,A. Zudov}
\email[Corresponding author: ]{zudov@physics.umn.edu}
\affiliation{School of Physics and Astronomy, University of Minnesota, Minneapolis, Minnesota 55455, USA}

\author{L.\,N. Pfeiffer}
\affiliation{Princeton University, Department of Electrical Engineering, Princeton, New Jersey 08544, USA}

\author{K.\,W. West}
\affiliation{Princeton University, Department of Electrical Engineering, Princeton, New Jersey 08544, USA}

\received{June 26, 2012}

\begin{abstract}
We have studied magnetotransport properties of a high-mobility two-dimensional electron system subject to weak electric fields.
At low magnetic field $B$, the differential resistivity acquires a correction $\delta  r \propto -\lambda^2 j^2/B^2$, where $\lambda$ is the Dingle factor and $j$ is the current density, in agreement with theoretical predictions.
At higher magnetic fields, however, $\delta r$ becomes $B$-independent, $\delta  r \propto -j^2$. 
While the observed change in behavior can be attributed to a crossover from overlapping to separated Landau levels, full understanding of this behavior remains a subject of future theories.
\end{abstract}
\pacs{73.43.Qt, 73.63.Hs, 73.21.-b, 73.40.-c}
\maketitle

Among many classes of magnetoresistance oscillations\citep{zudov:2001a,zudov:2001b,yang:2002,zhang:2007c,zhang:2008,khodas:2010,wiedmann:2010b} which occur in high Landau levels of two-dimensional electron systems (2DES), microwave-induced resistance oscillations (MIRO)\citep{zudov:2001a,ye:2001} are perhaps the best known and the most studied phenomenon, both theoretically\citep{durst:2003,lei:2003,vavilov:2004,dmitriev:2005,khodas:2008,dmitriev:2009b,hatke:2011e} and experimentally.\citep{mani:2002,zudov:2003,zudov:2004,willett:2004,mani:2004e,studenikin:2005,mani:2005,yang:2006,hatke:2008a,hatke:2008b,hatke:2009a,hatke:2011e} 
In the regime of overlapping Landau levels and low microwave power, theory predicts that high-order MIRO can be described by a radiation-induced correction to the resistivity (photoresistivity) of the form\citep{dmitriev:2009b}
\be
\frac {\delta \rho_\omega} {\rho} = - \aac \sin2\pi\eac\,,~~2\pi\eac\gg 1\,,\\
\label{eq.miro}
\ee
where $\rho$ is the resistivity without irradiation, $\eac=\omega/\oc$, $\omega=2\pi f$ and $\oc$ are the microwave and cyclotron frequencies,
\be
\aac=\aac_0 \lambda^{2}\,,~ \aac_0 = 4\pi\eac\pc\lp \frac {\ttr}{4\tst} + \frac {\tin}{\ttr} \rp\,,
\label{eq.aac}
\ee
$\lambda = \exp(-\pi/\oc\tq)$ is the Dingle factor, $\tq$ is the quantum lifetime, $\pc$ is the dimensionless microwave power,\citep{pechenezhskii:2007,khodas:2008} $\ttr$ is the transport lifetime, $\tst$ is the scattering time characterizing the correlation properties of the disorder potential,\cite{note:5} and $\tin$ is the inelastic relaxation time.
The first term in the parentheses in \req{eq.aac} accounts for the \emph{displacement} contribution,\citep{ryzhii:1970,ryzhii:1986,durst:2003,lei:2003,shi:2003,vavilov:2004,dmitriev:2009b} owing to the radiation-induced modification of impurity scattering, while the second term represents the \emph{inelastic} contribution,\citep{dorozhkin:2003,dmitriev:2003,dmitriev:2004,dmitriev:2005,dmitriev:2007,dmitriev:2009b} originating from the radiation-induced change in the electron distribution function.

Over the past decade, many experiments have examined the functional dependences of the MIRO amplitude $\aac$ on magnetic field $B$,\citep{zudov:2001a,hatke:2009a,hatke:2011b} microwave power $\pc$,\citep{ye:2001,zudov:2003,studenikin:2004,willett:2004,mani:2004a,mani:2010} and temperature $T$.\citep{studenikin:2005,studenikin:2007,hatke:2009a}
However, direct quantitative comparison of the measured MIRO amplitude to that predicted by \req{eq.aac} has not been attempted to date.
The main factor preventing such a study is an uncertainty in the microwave power $\pc$ absorbed by a 2DES.
Consequently, it is also not feasible to reliably evaluate the scattering parameters entering $\aac_0$ from the measured MIRO amplitude.
On the other hand, it is indeed very desirable to have a reliable experimental probe of such 2DES parameters as $\tst$ and $\tin$, which would allow characterization of the correlation properties of the disorder potential and the strength of interactions in a 2DES, respectively.

In this paper we propose and demonstrate an approach to experimentally evaluate $\tst$ and $\tin$ in high-mobility 2DES.
More specifically, we employ the nonlinear response of the resistivity to an applied dc field.
In contrast to studies investigating the regime of strong electric fields,\citep{yang:2002,bykov:2005c,zhang:2007a,zhang:2007c,zhang:2008,hatke:2008a,hatke:2008b,hatke:2011a} which is dominated by Hall field-induced resistance oscillations (HIRO),\citep{yang:2002} we focus on the regime of weak electric fields.
In this regime, to the second order in dc field, the theory\citep{vavilov:2007} predicts, in overlapping Landau levels, q dc-induced correction to the differential resistivity of the form
\be
\frac{\delta r_j}{\rho} = - \adc\edc^2\,, 
\label{eq.zbp}
\ee
where $\edc = \W/\hbar\oc$, $\W = 2\rc e \Edc$ is the work done by the electric field $\Edc$ over the cyclotron diameter $2\rc$, and
\be
\adc = \adc_0\lambda^{2}\,,~ \adc_0 = 12 \pi^2\lp \frac {3\ttr}{16\tst} + \frac {\tin}{\ttr} \rp\,.
\label{eq.adc}
\ee
Unlike the MIRO amplitude [\req{eq.aac}], which contains $\pc$, the curvature $\adc$ [\req{eq.adc}] contains only scattering parameters.

To examine the applicability of \reqs{eq.zbp}{eq.adc}, we have measured the differential resistivity in a high-mobility 2DES over a wide range of magnetic fields, covering the regimes of both overlapping and separated Landau levels.
At low magnetic fields, we have found that the differential resistivity acquires a correction which can be well described by \req{eq.zbp} with $\alpha \propto \lambda^2$, as prescribed by \req{eq.adc}.
The obtained value of $\alpha_0$ suggests that the response is dominated by the inelastic contribution given by the second term in \req{eq.adc}.
At higher magnetic fields, we observe a significant deviation from this behavior, which we attribute to a crossover from overlapping to separated Landau levels.
More specifically, at $B\gtrsim 1.3$ kG (1 kG = 0.1 T), we find $\alpha = 12\pi^2 (B/B_0)^2$, $B_0 \approx 0.93$ kG.
As a result, the correction to differential resistivity becomes \emph{independent} of $B$ and follows $\delta r/\rho =-j^2/j_0^2$, where $j_0 \approx 4.5\cdot 10^{-2}$ A/m.

Our Hall bar sample (width $w=100$ $\mu$m) was fabricated using photolithography from a symmetrically doped GaAs/Al$_{0.24}$Ga$_{0.76}$As 300\,\AA-wide quantum well grown by molecular beam epitaxy.
Ohmic contacts were made by evaporating Au/Ge/Ni, followed by rapid thermal annealing in forming gas.
The experiment was performed in a $^3$He cryostat, equipped with a superconducting solenoid, at a constant coolant temperature $T\simeq 1.5$ K.
After illumination with visible light, the electron density and mobility were \den and \mob, respectively.
The longitudinal differential resistivity $r=dV/dI$ was recorded using low-frequency (a few hertz) lock-in amplification as a function of $j=I/w$ at different fixed $B$ ranging between 0.3 and 2.2 kG.
The probing ac current was 0.2 $\mu$A.

\begin{figure}[t]
\includegraphics{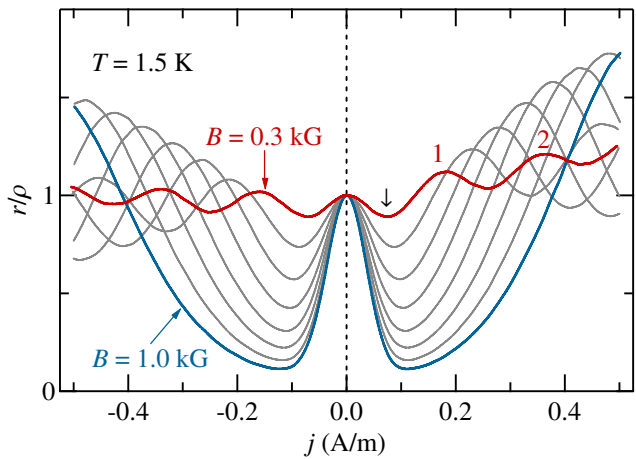}
\caption{(Color online) Normalized differential resistivity $r/\rho$ versus direct current of density $j$ at different magnetic fields from 0.3 kG (top curve) to 1.0 kG (bottom curve) in steps of 0.1 kG.
The maxima of HIRO measured at $B=0.3$ kG are marked by integers (cf. $1, 2$).
}
\label{fig1}
\end{figure}
In \rfig{fig1} we present the differential resistivity $r$, normalized to its value at zero current $\rho$, as a function of the current density $j$, measured at different magnetic fields from $B=0.3$ kG (top curve) to $B=1.0$ kG (bottom curve), in a step of 0.1 kG. 
The maxima of HIRO, which occur at $\edc = 1,2$, are marked by integers (cf.\,$1, 2$) next to the trace measured at $B=0.3$ kG.
With increasing $B$, these maxima shift to higher currents and eventually move outside the investigated current range.
The main focus of the present study, however, is the regime of small dc fields, $\edc \ll 1$, which, according to the theory,\citep{vavilov:2007} is described by \reqs{eq.zbp}{eq.adc}.
As seen from \rfig{fig1}, the nonlinearity in this regime becomes progressively stronger with increasing magnetic field.

\begin{figure}[t]
\includegraphics{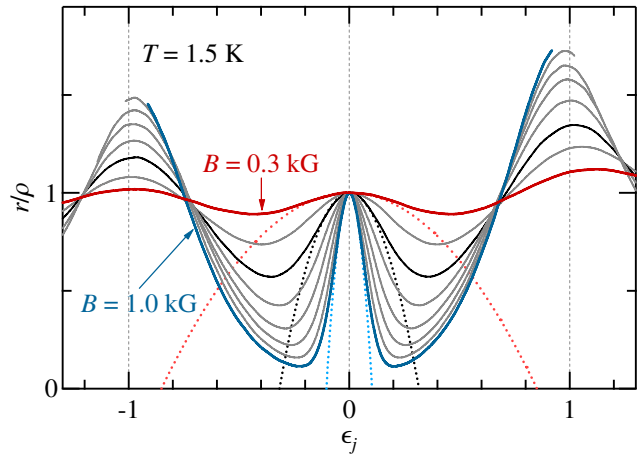}
\caption{(Color online) Normalized differential resistivity $r/\rho$ versus $\edc$ at different magnetic fields from 0.3 kG to 1.0 kG in steps of 0.1 kG (solid lines).
Dotted lines are fits to the data with $r/\rho = 1 - \adc \edc^2$ over the range $-0.1\le\edc\le 0.1$.
}
\label{fig2}
\end{figure}

Our goal is to analyze the data such as that shown in \rfig{fig1} in terms of \req{eq.zbp}, extract the curvature $\adc$, and then discuss it in the context of \req{eq.adc}.
After converting the current density $j$ to $\edc = 2 (2\pi/n_e)^{1/2} m^\star j/e^2B$,
where $m^\star\approx 0.067 m_{0}$ is the electron effective mass, we replot the data shown in \rfig{fig1} as a function of $\edc$ in \rfig{fig2}.
Presented this way, the differential resistivity shows the fundamental HIRO maxima at $\edc \approx \pm 1$ for \emph{all} magnetic fields, in agreement with previous experimental\citep{zhang:2007a,zhang:2007c,zhang:2008,hatke:2011a} and theoretical\citep{vavilov:2007,lei:2007,khodas:2008} studies. 
Our next step is to fit the data with $r/\rho = 1 - \adc \edc^2$ [cf. \req{eq.zbp}] over a range of low electric fields, $-0.1\le\edc\le 0.1$.
Three examples of such fits for $B=0.3$, 0.5, and 1.0 kG are shown in \rfig{fig2} by dotted lines.
It is clear that the curvature of the fits, $\adc$, grows rapidly with increasing $B$.

\begin{figure}[t]
\includegraphics{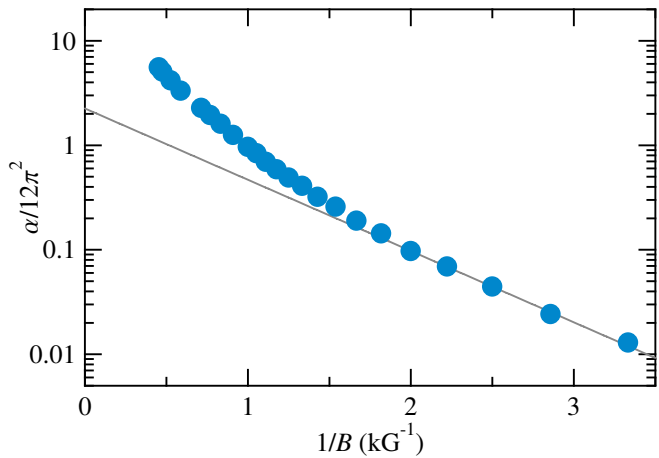}
\caption{(Color online) Obtained from the fits (cf. \rfig{fig2}) of parameter $\adc/12\pi^2$ (circles) versus inverse magnetic field $1/B$ plotted on a log-linear scale. 
The fit (solid line) to the lower $B$ part of the data, $B \le 0.5$ kG, with $\adc = \adc_0 \exp(-2\pi/\oc\tq)$--see \req{eq.adc}--generates $\tq = 15.2$ ps and $\adc_0/12\pi^2 = 2.25$.
}
\label{fig3}
\end{figure}

In \rfig{fig3} we present a parameter $\adc$ (circles), obtained from the fits to the data, such as that shown in \rfig{fig2}, versus inverse magnetic field $1/B$ plotted on a log-linear scale. 
Presented in such a way, the data reveal that the parameter $\adc$ changes by nearly three orders of magnitude over the studied $B$ range.
The lower $B$ ($B \le 0.5$ kG) part of the data can be well described by an exponential dependence, $\adc = \adc_0 \exp(-2\pi/\oc\tq)$, in accordance with \req{eq.adc}.
The slope of our fit to the data (solid line) generates $\tq = 15.2$ ps.
This value is close to $\tq$ obtained from the Dingle plots of MIRO and HIRO amplitudes, confirming the validity of our approach.
An estimate of $\adc_0$ is given by the intercept of the fit with the vertical axis.
We next analyze the value of $\adc_0/12\pi^2 = 2.25$, obtained from this intercept, in detail.

We first recall that the displacement contribution, given by $3\tst/16\ttr$, is sensitive to the correlation properties of disorder in the 2DES.
For example, for purely smooth disorder, the displacement contribution is the smallest, $\ttr/\tst  = 12/(\ttr/\tq-1) \approx 0.5$ ($3\ttr/16\tst \approx 0.09$).\cite{note:3}
In the opposite limit of only sharp disorder, $\ttr/\tst $ attains its maximal possible value, $\ttr/\tst = 3$.
We notice that even the maximal displacement contribution, $3\ttr/16\tst = 9/16 \approx 0.56$, is small compared to $\adc_0/12\pi^2 = 2.25$, obtained experimentally.
We thus must conclude that, regardless of the specifics of the disorder, the inelastic contribution dominates the nonlinear response resistivity in our high-mobility 2DES.
In lower mobility and higher density 2DES, the inelastic contribution becomes even stronger and the displacement contribution can be safely ignored, see, e.g. \rrefs{zhang:2007b}{zhang:2009}.
In our study, the ratio $\tin/\ttr$, which determines the inelastic contribution, is bounded by $1.69 \le \tin/\ttr \le 2.16$, from which we obtain $0.57~{\rm ns} \le \tin \le 0.73~{\rm ns}$.
This result agrees well with the theoretical estimate, $\tin \approx 0.56$ ns, obtained from $\hbar/\tin \simeq k_B^2 T^2/E_F$ ($E_F$ is the Fermi energy).\citep{dmitriev:2005}

We next obtain a more accurate estimate of the displacement and the inelastic contributions in our 2DES.
Using $\ttr/\tsh \approx 0.2$ ($\tsh^{-1}$ is the sharp disorder contribution to the quantum scattering rate) obtained from the $B$ dependence of the HIRO amplitude,\citep{hatke:2009c} we estimate $\ttr/\tst \simeq 3\ttr/\tsh + 12 \tq/\tau \approx 0.6 + 0.6 = 1.2$, a value which reflects approximately equal contributions from sharp and smooth components of disorder.\citep{dmitriev:2009b}
Using this estimate, we then obtain $3\ttr/16\tst \approx 0.23$, which leads to $\tin/\ttr \approx 2.0$ or $\tin \approx 0.64$ ns. 

The above analysis shows that the displacement mechanism contributes only a small fraction to the observed nonlinearity.
Since, theoretically, the relative contributions from the displacement and the inelastic mechanisms are essentially the same for both the MIRO amplitude, \req{eq.aac}, and nonlinear response resistivity, \req{eq.adc}, one should indeed expect that the displacement contribution to MIRO can also be neglected under similar experimental conditions.
However, a recent study examining the temperature dependence of the MIRO amplitude have found no $1/T^2$-dependence, characteristic of the inelastic mechanism.\citep{hatke:2009a} 
This apparent controversy can be, at least partially, resolved by noticing that the density (mobility) of the 2DES used in \rref{hatke:2009a} was lower (higher) compared to that of the 2DES investigated here.
As a result, the inelastic contribution, which scales with $\tin/\ttr \propto n_e/\mu$, was at least twice as small compared to the present study.
We also note that under experimental conditions of \rref{hatke:2009a}, the temperature dependence of the MIRO amplitude $A=\lambda^2 A_0$ was dominated by an exponentially changing $\lambda^2$, which significantly complicates detecting the temperature dependence of $\tin/\ttr$ entering $A_0$.
To confirm the inelastic contribution in microwave photoresistance, it is very desirable to investigate the MIRO temperature dependence at lower temperatures where $\lambda^2$ becomes $T$ independent.
Such a study, however, is complicated by a radiation-induced heating of the 2DES, which gets progressively stronger at lower temperatures.

\begin{figure}[b]
\includegraphics{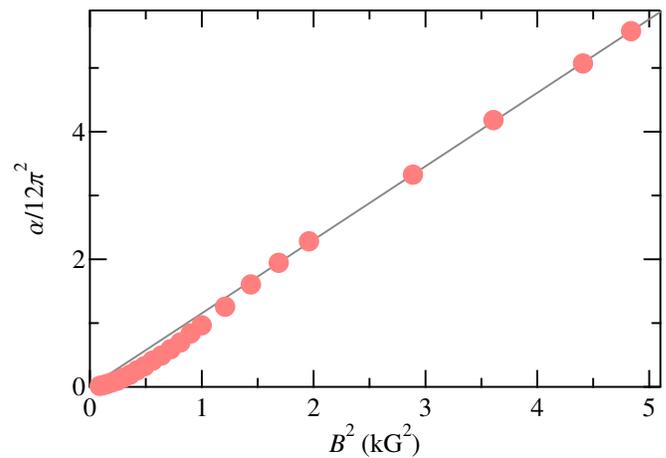}
\caption{(Color online) Obtained from the fits (cf. \rfig{fig2}) parameter $\adc/12\pi^2$ (circles) versus $B^2$. 
The fit (solid line) to the higher $B$  part of the data, $B \ge 1.3$ kG, with $\adc/12 \pi^2 =  B^2/B_0^2$ generates $B_0 \approx 0.93$ kG.
}
\label{fig4}
\end{figure}

Further examination of \rfig{fig3} shows that, at higher magnetic fields, $\adc$ grows faster than the exponential dependence predicted by \req{eq.adc}.
A departure from the exponential behavior is likely a signature of a crossover between the regimes of overlapping and separated Landau levels. 
Indeed, using the condition $\oc\tq = \pi/2$,\citep{ando:1974b,laikhtman:1994} we find that the Landau levels separate at a magnetic field of $\approx 0.4$ kG.
Examination of \rfig{fig3} confirms that the magnetic field, at which the departure from the exponential dependence occurs, compares well with this estimate.

To further examine the regime of separated Landau levels, we replot $\adc$ (circles) in \rfig{fig4} as a function of $B^2$.
The fit (solid line) to the higher $B$  part of the data, $B \ge 1.3$ kG, shows that the data in this regime can be well described by $\adc/12 \pi^2 =  B^2/B_0^2$ with $B_0 \approx 0.93$ kG.
Since $\edc \propto 1/B$, this observation suggests that the correction to the differential resistivity becomes $B$ independent and is determined \emph{only} by the applied current $j$.
Indeed, one can write $\delta r / \rho = - j^2/j_0^2$, where $j_0 = e^2 B_0 \sqrt{n_e}/4\pi\sqrt{6\pi}m^\star\approx 4.5 \cdot 10^{-2}$ A/m in our 2DES.

Within a framework of the displacement mechanism, such a behavior can be qualitatively understood by noting that at low electric fields and in separated Landau levels, the nonlinear response of the 2DES is governed by impurity scattering within a single Landau level, which is located near the Fermi surface.
In this situation, the inter-Landau level spacing is no longer important, and the relevant energy scale is given by the Landau level width $\Gamma$. 
As a result, $\edc=\W/\hbar\oc$ in \req{eq.zbp} should be replaced by $\beta \W/ \Gamma$, where $\beta$ is a constant of the order of unity, which depends on the functional form of the density of states.
The correction to the differential resistivity then takes a form $\delta r_j/\rho \propto - \W^2/\Gamma^2$.
Since $\W \propto j$ and is $B$ independent, one obtains $\delta r_j/\rho \propto - j^2/j_0^2$.
Therefore, our finding that $j_0$ does not depend on $B$ implies a $B$-independent $\Gamma$.
While a theoretical expression for $\delta r$ in separated Landau levels is not currently available, we note that $\W(j_0) = 2\hbar\sqrt{2\pi n_e}j_0/e \approx 0.17$ K compares well with $\hbar/2\tq \approx 0.25$ K.

\begin{figure}[t]
\includegraphics{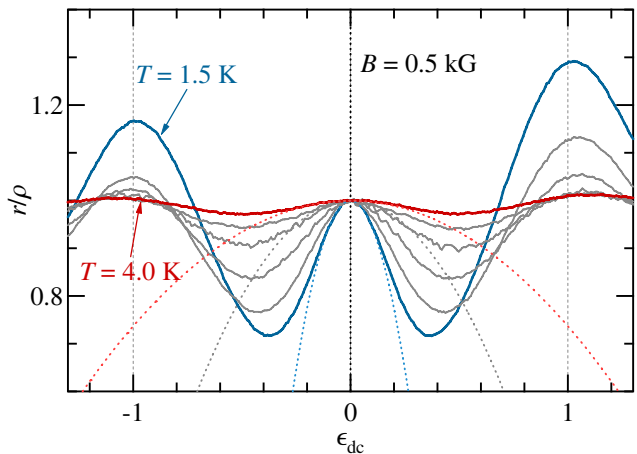}
\caption{(Color online) 
Normalized differential resistivity $r/\rho$ versus $\edc$ measured at $B = 0.5$ kG at $T$ from 1.5 K to 4.0 K, in a step of 0.5 K (solid lines).
Dashed lines are fits to the data with $r/\rho = 1 - \adc \edc^2$ over the range $-0.1 \le \edc \le 0.1$.
}
\label{fig5}
\end{figure}

Finally, we note that the observed nonlinearity weakens considerably with increasing temperature.
In \rfig{fig5} we present normalized differential resistivity $r/\rho$ versus $\edc$ measured at $B = 0.5$ kG  at $T$ from 1.5 K to 4.0 K, in a step of 0.5 K (solid lines).
The fits with $r/\rho = 1 - \adc \edc^2$ (dotted lines for 1.5, 3.0, and 4.0 K) demonstrate that $\delta r$ can still be described by \req{eq.zbp} for all temperatures studied and that the curvature $\alpha$ decays rapidly with increasing $T$.
The main source of this decay is the increase of the electron-electron scattering, $1/\tin \propto T^2$.
This scattering not only suppresses the inelastic contribution, given by the second term in \req{eq.adc}, but also modifies the quantum scattering rate, $1/\tq$.\citep{hatke:2009a,hatke:2009b,hatke:2009c}
The latter results in the suppression of both the displacement and the inelastic contributions, since both scale with $\lambda^2=\exp(-2\pi/\oc\tq)$, see \req{eq.adc}.
Another source of temperature dependence is the enhanced scattering on thermal acoustic phonons, which modifies the transport scattering rate, $1/\ttr$, and gives rise to phonon-induced resistance oscillations.\citep{zudov:2001b,hatke:2009b,hatke:2011d}
These oscillations are known to interfere with the nonlinear response resistivity,\citep{zhang:2008,dmitriev:2010b} resulting in a nontrivial, $B$ dependent corrections to $\alpha$ in \req{eq.zbp}.
Finally, we mention a recently reported negative magnetoresistivity effect\citep{bockhorn:2011,hatke:2012a} which occurs in the same range of magnetic fields and is strongly temperature dependent.
Unfortunately, separating all these contributions does not appear feasible at this point.

In summary, we have studied the nonlinear resistivity of a high-mobility 2DES over a range of magnetic fields covering the regimes of both overlapping and separated Landau levels.
At low magnetic fields, we have found that the differential resistivity acquires a correction which can be well described by $\delta r \propto - \exp(-2\pi/\oc\tq) j^2/\oc^2$.
Quantitative comparison with existing theory\citep{vavilov:2007} indicates that the nonlinear response in our 2DES is dominated by the inelastic contribution.
At higher magnetic fields, we observe a significant deviation from the above exponential dependence, which we attribute to the crossover from overlapping to separated Landau levels.
Here, the correction to differential resistivity becomes \emph{independent} of $B$ and can be well described by $\delta r/\rho = -j^2/j_0^2$, where $j_0 \approx 4.5\cdot 10^{-2}$ A/m.
It will be interesting to see if future theories can explain this finding and clarify the physical meaning of $B$ independent $j_0$.

We thank I. Dmitriev and T. Shahbazyan for discussions.
The work at Minnesota was supported by US Department of Energy, Office of Basic Energy Sciences, under Grant No. DE-SC002567. 
The work at Princeton was partially funded by the Gordon and Betty Moore Foundation and by the NSF MRSEC Program through the Princeton Center for Complex Materials (DMR-0819860).


\begin{thebibliography}{56}
\bibitem[{not({\natexlab{a}})}]{note:5}
\bibinfo{note}{The rate of scattering on angle $\theta$ can be expressed in
  terms of angular harmonics, $\tau_n=\tau_{-n}$, as $\tau_\theta^{-1} =
  \sum\tau_n^{-1}e^{i n\theta}$. In this notation, $\tq^{-1}=\tau_0^{-1}$,
  $\ttr^{-1}=\tau_0^{-1}-\tau_1^{-1}$, and
  $\tst^{-1}=3\tau_0^{-1}-4\tau_1^{-1}+\tau_2^{-1}$.}
\bibitem[{not({\natexlab{b}})}]{note:3}
\bibinfo{note}{In our 2DES, the transport scattering time is $\ttr = \mu m^\star/e
  \approx 0.34$ ns.}
\end{thebibliography}
\end{document}